\documentclass{ws-p8-50x6-00}

\def\NPPS{\em Nucl. Phys. Proc. Suppl.}
\def\ANP{\em Annals of Phys.}

\def\la{\langle} 
\def\ra{\rangle} 
\def\be{\begin{eqnarray}} 
\def\ee{\end{eqnarray}}

\newcommand{\eq}{\begin{equation}} \newcommand{\eqx}{\end{equation}}

\newcommand{\eqn}{\begin{eqnarray}} \newcommand{\eqnx}{\end{eqnarray}}

\renewcommand{\th}{\theta}
\newcommand{\sg}{\sigma}

\newcommand{\chis}{\chi_*}

\newcommand{\nn}{\mbox{\bf{}n}}

\begin{document}

\title{$\theta$ Vacuum in a Random Matrix Model}

\author{Romuald A. Janik$^{1,2}$, Maciej A.  Nowak$^{2,3}$, 
	G\'abor Papp$^{4}$ \\ \lowercase{and} Ismail Zahed$^5$}
\address{$^1$ Service de Physique Th\'{e}orique, CEA Saclay, 
F-91191 Gif-Sur-Yvette, France.\\
$^2$ Department of Physics, Jagellonian University, 30-059
Krakow, Poland.\\ $^3$ GSI, Planckstr. 1, D-64291 Darmstadt, Germany\\ 
$^4$CNR Department of Physics, KSU, Kent, Ohio 44242, USA \& \\
HAS Research Group for Theoretical Physics, E\"{o}tv\"{o}s University, 
Budapest,\\ 
$^5$Department of Physics and Astronomy,
 SUNY, Stony Brook, New York 11794.}
\maketitle

\abstracts{%
Inspired by recent lattice calculations, we model certain aspects of the
$\theta$-vacuum using a matrix model with gaussian weights. The vacuum 
energy exhibits a cusp at $\theta <\pi$ that is sensitive to both the
accuracy of the numerical analysis and the maximum density of
winding modes present in a finite volume.}

Recent lattice study of the $CP^3$ model on the lattice\cite{SCHIE} have
revealed an intriguing result: the energy as a function of the CP
breaking $\th$ angle levels off at large $\th$.
The result is very interesting, since in that model the string 
tension for small charges is related to the $\th$-derivative of the 
vacuum energy. A plateau indicates the loss of confinement.
The result was reexamined in~\cite{SAM} where the authors pointed
out to some subtleties regarding the technical methods applied.

In this talk we will address these issues using a random matrix model 
with the partition function
\be
  Z(\theta, N_f)=\left\la\prod_{j=1}^{N_f}
                 {\rm det}{\left(\begin{array}{cc}
	im_je^{i\theta/N_f}        & W \cr
	W^\dagger & im_je^{-i\theta/N_f}
  \end{array}\right)}\right\ra \,.
\label{partsumf}
\ee
where the averaging is carried over the matrices $W$, $W^\dagger$ using
the weight $\exp{(-n/2 {\rm Tr} W^\dagger W)}$ and over the matrix size $n$
and $\sg$, with weights $\exp{(-{\chi}^2/{2\chi_* V})}$
and $\exp{(-\sigma^2/{2\sigma_*V})}$, respectively.
Here $W$ is a complex asymmetric $n_+\times n_-$ matrix,
$n=n_++n_-$, and $\sigma\pm \chi =2n_{\pm} - \la n\ra$. 
The mean number of zero modes $\la n\ra$ is either fixed from the outside 
or evaluated using the gaussian measure. Here and for simplicity we
use the quenched measure without the fermion determinant to fix $\la n\ra$.
Throughout, we set the value of the quark condensate to $\Sigma=1$ in the 
chiral limit.

Eq.~(\ref{partsumf}) is borrowed from the effective instanton
vacuum analysis~\cite{US} where $n_+$ counts the number of right-handed
zero modes , and $n_-$ the number of left-handed zero modes. The number
of exact topological zero modes is commensurate with the net winding number 
carried by the instantons and antiinstantons.  By analogy with~\cite{US}, 
$\chi_*$ and $\sigma_*$ will refer to the unquenched topological 
susceptibility and particle compressibility, respectively, with  
$\sigma_*^2=12\nn_*/11N_c$ and $\nn_* =\la n\ra/V$ the mean density of 
zero modes. If the compressibility $\sigma_*$ is assumed small in units of 
$\Sigma =1$, then typically $n\sim \la n\ra$.

\vspace*{1mm}
{\bf quenched case: }
In the case of the quenched partition function ($N_f=0$) one 
probes the nature of the gaussian measure. We start discussing
the case where $\la n\ra=\infty$ with no restriction
on the value of $n$, and hence no restriction on the value of 
$\chi$. Then, we discuss the case where $\la n\ra$ is large but finite, 
so that $|\chi | \leq n$ with typically $n\sim \la n\ra$ for a peaked
distribution in $n$.

When the sum is unrestricted and infinite, using Poisson resummation formula
we have
\be
Z_Q(\theta ) = \sum_{k=-\infty}^{+\infty}
e^{-\frac 12 V \chi_* ({\theta} - 2\pi k)^2}
=\theta_3 (\theta/2, e^{-\tau}) \,.
\label{POISSON}
\ee
with $\theta$ being the third elliptic function and
$\tau=1/(2V\chi_*)$. The result is manifestly $2\pi$ periodic. The
vacuum  energy, $F_Q(\theta) =-{\rm ln}Z_Q (\theta)/V$ as 
$V\rightarrow\infty$ is simply
$
F_Q(\theta ) = {\rm min} \frac 12 \chi_* \left(\theta  +{\rm mod}\, 2\pi
\right)^2
$
in agreement with the saddle-point approximation.
This result is in agreement with the result using large 
$N_c$ arguments~\cite{WITTEN}, and  recent
duality arguments~\cite{WITTENFRESH}. 
We observe that the cusp at $\theta=\pi$ (mod $2\pi$)
sets in for $V=\infty$. 

\begin{figure}[htbp]
\centerline{\epsfysize=40mm \epsfbox{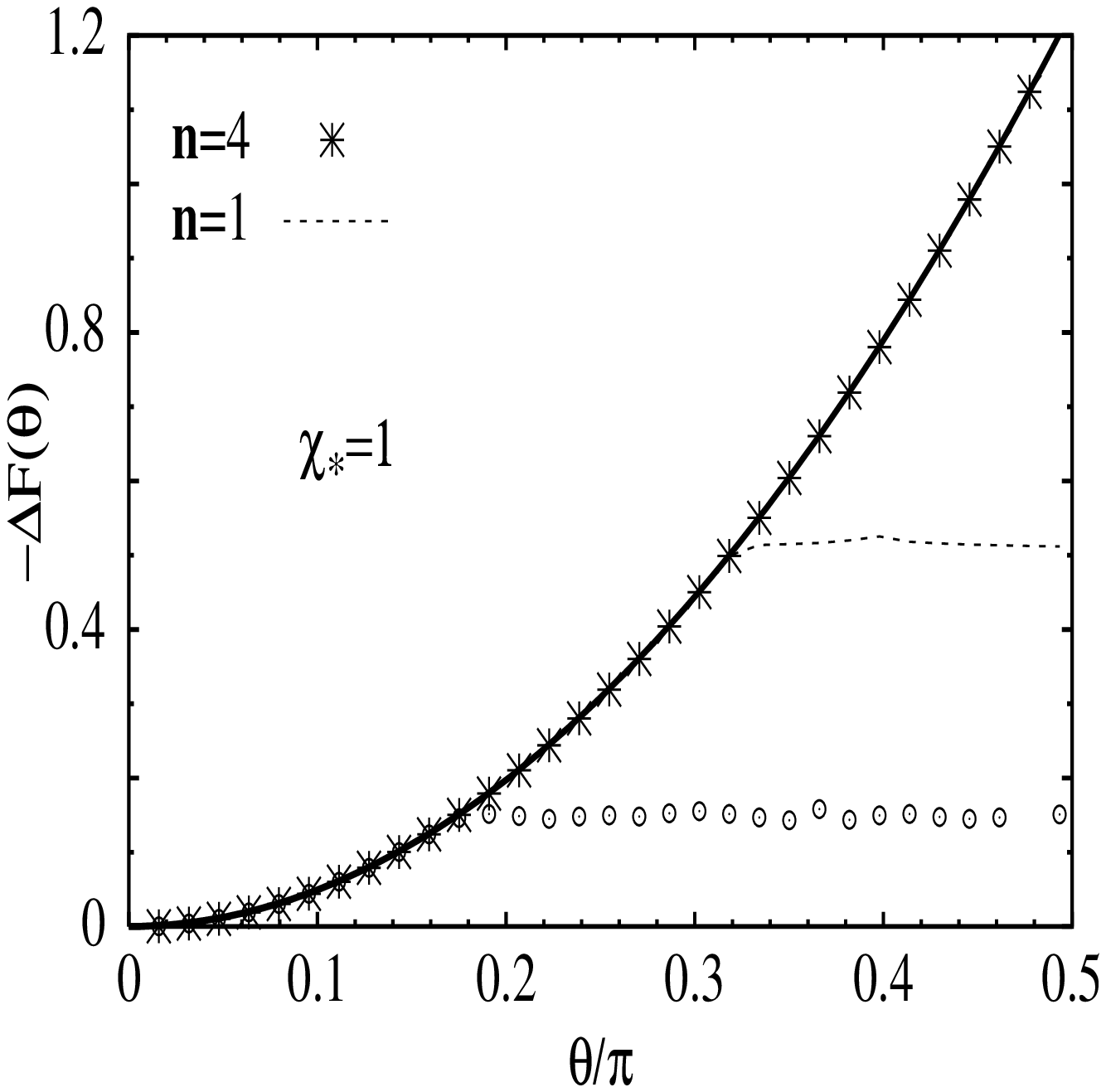} \hspace*{5mm}
\epsfysize=40mm \epsfbox{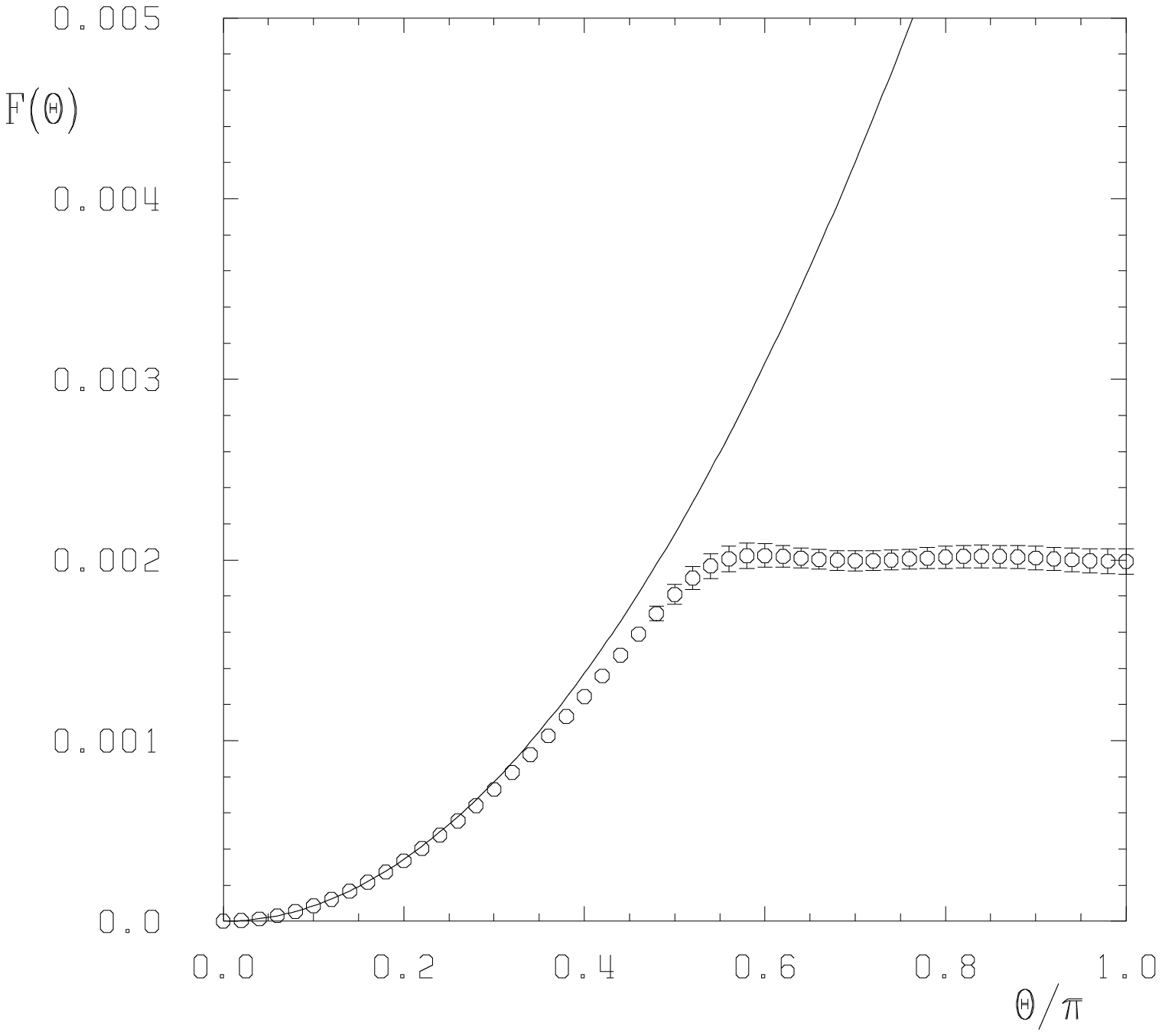}}
\vspace*{-3mm}
\caption{Free energy in a quenched model (left) and in
the $CP^3$ model (right). See text.}
\label{fig-que}
\end{figure}

In the matrix model being considered the sum over $\chi$ is
restricted to $|\chi |< N$, with $N={\rm max}\,\, n$. We denote by 
$\nn =N/V$ the maximum density of winding modes. While in general
$\nn\neq \nn_*$, for a peaked distribution in $n$ (small 
compressibility $\sigma_*$) we expect $\nn\sim \nn_*$. This 
will be assumed throughout unless indicated otherwise. Hence
\be
Z_Q (\theta) = \sum_{\chi=-(N-1)}^{N-1} e^{i\theta\chi} e^{-\chi^2/2V\chi_*}
\label{que-num}
\ee
Approximating the sum in (\ref{que-num}) by an integral and evaluating
it by saddle point we obtain $Z_Q \sim e^{-V\chis\th^2/2}$ apparently in
agreement with the infinite sum. However, the Euler-MacLaurin summation
formula shows that deviation from this result in the case of a finite sum
is expected for $\th>\th_c\sim\nn/\chis$~\cite{USTHETA}. Indeed, this is
confirmed by detailed numerical calculations of the vacuum energy as shown in
Fig~\ref{fig-que} (left).  For $N=250$ and $\chi_*=1$, the double
precision (16 digit) numerics (circles) breaks away from the saddle
point approximation (solid line) at $\theta/\pi\sim 0.2$ for both
$\nn=1$ and $\nn=4$, while the high-precision (64 digits)
calculations (dashed line) agree with the saddle point result at $\nn=4$
but break away at $\theta/\pi\sim 0.3\approx 1/\pi\chis$ at $\nn=1$. On
the right we show the numerical result for the $CP^3$ model~\cite{SCHIE}
showing a similar behavior.

\vspace*{-4mm}
{\bf unquenched case: } Analytical and numerical studies of the unquenched 
matrix model for $N_f=1$ and $N_f=2$~\cite{USTHETA} lead to results similar to
the ones observed in the quenched case. The mean-field result breaks down 
for $\th>\th_c\sim \nn/\chi_{top}$. For
small quark masses $\chi_{top}\sim m$~\cite{U1US}, the breakdown
is observed only for ensembles with very small maximal winding
number.

In this talk we have shown that the vacuum energy of a chiral matrix model
developes a cusp at $\theta<\pi$ that is sensitive to the numerical accuracy.
The cusp persits at high accuracy if the maximum winding number considered
is low, in disagreement with mean-field results. Similar observations in 
current lattice simulations should therefore be interpreted with care.

\vspace*{-3mm}
\section*{Acknowledgments}
\vspace*{-2mm}

This work was supported by the US DOE grants DE-FG-88ER40388,
by the Polish Government Project (KBN)  grant 
2P03B00814 and by the Hungarian grant OTKA-F026622.


\vspace*{-3mm}
\begin{thebibliography}{50}
\vspace*{-2mm}
\bibitem{SCHIE} G. Schierholz,
        \Journal{\NPPS}{A37}{1994}{203}.

\bibitem{SAM} J. C. Plefka and  S. Samuel,
        \Journal{\PRD}{56}{1977}{44}.


\bibitem{US}
M. Nowak, J. Verbaarschot and I. Zahed,
        \Journal{\PLB}{228}{1989}{251}.

\bibitem{WITTEN}
E. Witten,
        \Journal{\ANP}{128}{1980}{363}.

\bibitem{WITTENFRESH}
E. Witten, 
        \Journal{\PRL}{81}{1998}{2862}.

\bibitem{USTHETA}
R.A. Janik, M.A. Nowak, G. Papp and I. Zahed,
        hep-ph/9901390.

\bibitem{U1US}
R.A. Janik, M.A. Nowak, G. Papp and I. Zahed,
        \Journal{\NPB}{498}{1997}{313}.

\end{thebibliography}
\end{document}